\documentclass{jpsj3}
\usepackage{subfigure}
\usepackage{amsmath}
\usepackage{amssymb}
\title{Stability in Microcanonical Many-body Spin Glasses}
\author{\name{Zsolt \surname{Bertalan}}
\thanks{Email address: zsolt@stat.phys.titech.ac.jp} and 
\name{Kazutaka \surname{Takahashi}}}

\newcommand{\be}{\begin{eqnarray}}
\newcommand{\ee}{\end{eqnarray}}
\newcommand{\no}{\nonumber}
\newcommand{\salb}{\sum_{\alpha<\beta}}

\newcommand{\ab}{_{\alpha\beta}}
\newcommand{\al}{_{\alpha}}
\newcommand{\sr}{_{\sigma\rho}}
\newcommand{\sslr}{\sum_{\sigma<\rho}}
\newcommand{\ssr}{\sum_{\sigma\rho}}

\kword{spin glasses, ensemble inequivalence, stability, 
microcanonical ensemble, phase diagram}

\abst{We generalize the de Almeida-Thouless line for the many-body 
Ising spin glass to the microcanonical ensemble and show that 
it coincides with the canonical one. 
This enables us to draw a complete microcanonical phase diagram of this model.}

\inst{\address{Department of Physics, Tokyo Institute of Technology, 
Tokyo 152-8551, Japan.}}

\begin{document}
\maketitle
\section{Introduction}
Spin glasses are magnetic systems in which 
spin interactions take random values and 
remain fixed while the value of the spins may fluctuate. 
This is called quenched randomness. 
In such systems there occurs under certain conditions 
a thermodynamic phase transition where spins are frozen 
in random orientations, and thus their overall configuration 
does not change, the spin-glass phase. 

To calculate physical observables under quenched randomness 
we average over the random interactions. 
This is possible since extensive quantities, like the free energy, 
are self-averaging which means that they are the same for any realization 
of the random variables. 
This usually entails averaging over complicated expressions, since 
the free energy for example, is the logarithm of the partition function. 
To deal with these difficulties, the replica trick was invented \cite{EA}. 
It relies on the simple equation: 
$\ln x = \lim\limits_{n\to 0} (x^n-1)/n$. 
The notion behind this formalism is that it is much easier to evaluate 
some power of the partition function than its logarithm. 

In the theory of spin glasses, 
the de Almeida-Thouless (AT)-line \cite{AT} is a boundary below which 
the replica symmetric solution of the Sherrington-Kirkpatrick (SK) 
model \cite{SK} becomes unstable. 
It was generalized later by Gardner to include many-body spin glasses 
and replica-symmetry breaking \cite{gardner}. 
We established in a recent paper that for many-body spin glasses, 
the boundary between the paramagnetic and ferromagnetic phases 
is dependent on whether it was derived in the canonical or 
microcanonical ensemble \cite{zbhn}. 
There is, however, no ensemble inequivalence where 
the spin-glass phase is concerned. 
It is therefore expected that the result for the AT line 
there is the same in both ensembles. 
In this paper we prove this assumption. 

Ensemble inequivalence occurs in systems with long-range interactions 
which are not additive. 
That means that when two subsystems with energy $E$ are brought into 
contact their total energy will not be $2E$ in general. 
Many, at first glance, counter-intuitive effects appear in 
the microcanonical treatment of long-range interacting systems, 
like the appearance of negative specific heat. 
For a review of ensemble inequivalence, see e.g. Campa et al.~\cite{campa}. 

The second purpose of this article is to draw 
the complete microcanonical phase diagram, 
including the AT line, minimal attainable energy, 
and the Nishimori line (NL) \cite{nml1,nml2} 
of the many-body Ising spin glass.

This paper is structured as follows. 
Following this introduction, in sec.~\ref{model} 
we introduce briefly the model.
In sec.~\ref{ATL}, we derive the microcanonical AT line, 
and discuss the condition of the NL in sec.~\ref{NL}. 
In sec.~\ref{PD}, 
we draw the complete phase diagram of the three-body spin glass. 
Paragraph~\ref{CC} is devoted to concluding remarks.

\section{Model}\label{model}

We restrict ourselves to the analysis of the Ising model 
with $p$-body, infinite-range interactions, 
called the infinite-range model. 
It is given by the Hamiltonian 
\be
 H=-\sum_{i_1<..<i_p}J_{i_1..i_p}S_{i_1}...S_{i_p}
 \quad\quad \mbox{with}\qquad S_{i_k}=\pm 1,
\ee
where the bonds are random numbers drawn independently from the distribution
\be
 P(J_{i_1..i_p})=\left(\dfrac{N^{p-1}}{\pi p!}\right)^{1/2}
 \exp\left\{-\dfrac{N^{p-1}}{p!}
 \left(J_{i_1..i_p}-\dfrac{j_0p!}{N^{p-1}}\right)^2\right\},
 \label{prob}
\ee
with mean $j_0p!/N^{p-1}$, and $j_0$ is called the ferromagnetic bias.

Using the replica trick, the microcanonical entropy density $s$, 
times the number of replicas $n$, is easily obtained \cite{zbhn}. 
It is written with respect to spin-glass order parameter  
$q_{\alpha\beta}\sim \sum_i S_i^{\alpha}S_i^{\beta}/N$ 
and magnetization $m_{\alpha}\sim \sum_i S_i^{\alpha}/N$ as 
\be\label{S}
 & & ns= -\sum_{\alpha\beta}\epsilon_{\alpha}
 (Q^{-1})_{\alpha\beta}\epsilon_{\beta}
 -\sum_{\alpha <\beta}q_{\alpha\beta}\hat q_{\alpha\beta}
 -\sum_{\alpha} m_{\alpha}\hat m_{\alpha}+\ln \Tr e^L, 
 \no\\
 & & L=\sum_{\alpha > \beta}\hat q_{\alpha\beta} S^{\alpha} S^{\beta}
 +\sum_{\alpha} \hat m_{\alpha} S^{\alpha}
\ee
where $\epsilon_{\alpha}=\epsilon+j_0m_{\alpha}^p$, 
$Q_{\alpha\beta}=\delta\ab+q^p\ab$, and $q_{\alpha\alpha}=0$.
The parameters $\hat q\ab$ and $\hat m_{\alpha}$ are obtained from 
the saddle point conditions $\partial s/\partial q\ab=0$ and 
$\partial s/\partial m_{\alpha}=0$ as
\be\label{spc}
 & & \hat q\ab = 2pq\ab ^{p-1}\ssr \epsilon\al 
 Q^{-1}_{\alpha\sigma}Q^{-1}_{\rho\beta}\epsilon_{\beta}, \\
 & & \hat m\al =2pm\al ^{p-1}j_0\sum_{\beta}Q^{-1}_{\alpha\beta}\epsilon_{\beta}.
\ee

\section{Microcanonical AT Line}\label{ATL}

In this section we derive the microcanonical AT line 
for the replica symmetric (RS) and 
first-step replica symmetry breaking (1RSB) cases.

\subsection{General Considerations}

The entropy as given in eq.~\ref{S} cannot be solved in full generality. 
Usually, one makes some allowance for the symmetry in the replicas. 
For example, in the replica symmetric ansatz, 
one assumes that all replicas are identical and share the same values 
of order parameters. 
As it turns out, this ansatz leads to negative entropy 
at some finite temperature and is thus unphysical. 
However, it is valid at higher temperatures. 
Similarly, replica-symmetry braking ansatz may yield negative entropies 
at low temperatures. 

Let us now assume that we imposed some symmetry on the replicas 
and label the appearing parameters and quantities with a zero, i.e. 
the entropy $s\to s_0$ or the spin-glass order parameter $q\ab^0$. 
To investigate the stability of this ansatz, 
we expand the entropy around the imposed solution in fluctuations 
($q\ab\to q^0\ab+dq\ab$, $m\al\to m^0\al+dm\al$) to second order
$s\to s_0+\sum dq\ab A_{\alpha\beta\gamma\delta}dq_{\gamma\delta}+\ldots$.
Since we used the saddle-point method to obtain the entropy 
from the sum of states, 
we require that the Hessian $A$ has only positive eigenvalues. 
If this is violated at some point, 
then the imposed symmetry is obviously no longer valid.

Calculation of the eigenvalues of the Hessian 
yields that there is indeed an eigenvalue, 
called the replicon mode, which may become negative at some temperature. 
However, it does not involve $fluctuations$ of the magnetization. 
A careful examination of the expansion of the microcanonical entropy 
allows us to draw parallels to the canonical case and 
conclude that the microcanonical replicon mode will also be independent 
of fluctuations of $m$. 
Therefore, we set subsequently $dm\al=0$, 
which also implies $d\hat m\al=0$. 

Using the matrix expansion of the inverse of $Q$
\be
 Q^{-1}\approx(Q_{0}+dQ+d^2Q)^{-1}\approx 
 Q^{-1}_{0}-Q^{-1}_{0}(dQ+d^2Q)Q^{-1}_{0}+Q^{-1}_{0} dQ Q^{-1}_{0} dQ Q^{-1}_{0},
\ee
and the expansion of the Fourier mode of spin-glass 
order parameter $\hat q\ab\to \hat q^0\ab+d\hat q\ab + d^2\hat q\ab$, where
\be
 d\hat q\ab &=& 
 d\left( 2pq\ab ^{p-1}\ssr \epsilon\al 
 Q^{-1}_{\alpha\sigma}Q^{-1}_{\rho\beta}\epsilon_{\beta}\right) \no\\
 &\approx& 2p(p-1)q^{p-2}\ab\epsilon^2 t^2 dq\ab - 2pq^{p-1}\ab 
 \epsilon^2 t^2\sum_{\sigma\rho}\left((Q^{-1}_{0})_{\alpha\sigma}
 +(Q^{-1}_{0})_{\beta\rho}\right) pq^{p-1}\sr dq\sr,
\label{TM}
\ee
with $t=\sum_{\alpha}(Q^{-1}_{0})_{\alpha\beta}$, 
the expansion of the entropy in eq.~(\ref{S}) can be calculated 
in a straightforward manner. 
It is given by
\be\label{s_2}
 ns &\approx& ns_0 -\dfrac{1}{2}\salb dq\ab d\hat q\ab \no\\
 & & +\dfrac{1}{2}\salb\sslr d\hat q\ab d\hat q\sr 
 (\langle S^{\alpha}S^{\beta}S^{\sigma}S^{\rho}\rangle
 -\langle S^{\alpha}S^{\beta}\rangle\langle S^{\sigma}S^{\rho}\rangle).
\ee
The average over the spins is the weighted trace, 
$\langle\cdots\rangle=\Tr (\cdots) e ^{L_0}$, where 
\be
L_0=\salb\hat q\ab^0  S^{\alpha}S^{\beta}+\sum_{\alpha} \hat m_{\alpha}^0 S^{\alpha}.
\ee
We have already dropped the terms linear in fluctuations 
since the entropy must be extremized with respect to all 
its variables and thus, those terms disappear.

If we take the $n(n-1)/2$ distinct elements of $dq\ab$ 
and arrange them as a vector $dq_i$, we can express the entropy as
\be
 ns=ns_0-\dfrac{1}{2}\sum_{ij}^{n(n-1)/2}dq_i A_{ij}dq_j, 
\ee
where the Hessian is made up of two matrices $A=T^{-1}-G$. 
Here, $T$ is defined as the relation between $dq\ab$ and $d\hat q\ab$ 
as given in eq.~(\ref{TM}), i.e. $d\hat{q}_i=\sum_jT_{ij}dq_{j}$, 
and the matrix $G$ is implicitly defined in eq.~(\ref{s_2}).

The calculation of the eigenvalues of $A$ is somewhat tedious and 
we refer to the literature \cite{nishimoribook} for details. 
It is sufficient for our purposes to note that the replicon mode 
can be expressed as
\be
 \lambda_3= (\hat P - 2\hat K + \hat R)^{-1} + (P-2K+R),
\ee
where
\be
 && \hat P =2 p(p-1)\epsilon^2 t^2 q^{p-2}
 -2p^2\epsilon^2 t^2 q^{2p-2}2((Q_0)^{-1}_{\alpha\alpha}+(Q_0)^{-1}_{\alpha\beta}), \\
 && \hat K  = -2p^2\epsilon^2 t^2 q^{2p-2}2((Q_0)^{-1}_{\alpha\alpha}
 +3(Q_0)^{-1}_{\alpha\beta}), \\
 && \hat R  = -2p^2\epsilon^2 t^2 q^{2p-2}2(Q_0)^{-1}_{\alpha\beta},
\ee
are elements of the matrix $T$ and 
\be
 && P=1-\langle S^{\alpha}S^{\beta}\rangle\langle S^{\sigma}S^{\rho}\rangle,\\
 && K=\langle S^{\alpha}S^{\beta}\rangle
 -\langle S^{\alpha}S^{\beta}\rangle\langle S^{\sigma}S^{\rho}\rangle, \\
 && R=\langle S^{\alpha}S^{\beta} S^{\sigma}S^{\rho}\rangle
 -\langle S^{\alpha}S^{\beta}\rangle\langle S^{\sigma}S^{\rho}\rangle,
\ee
are elements of $G$.

\subsection{Replica Symmetry}

In the RS ansatz it is assumed that the values of the order parameters 
are equal in all replicas, 
$m\al=m$ and $q\ab=q$, for any $\alpha,\beta$. 
In this formulation, we have $Q^{0}\ab=q^p$ for $\alpha\neq\beta$ 
and the components of $\lambda_3$ are easily listed as
\be
 \hat P -2\hat K +\hat R 
 &=& \frac{2p(p-1)q^{p-2}(\epsilon+j_0m^p)}{(1-q^p)}, \\
 P-2K+R &=& 1-q^2-2(q-q^2)
 +\langle S^{\alpha}S^{\beta}S^{\sigma}S^{\rho}\rangle-q^2 \no\\
 &=& \int Du  \cosh^{-4}(\sqrt{\hat q}u + \hat m),
\ee 
where $Du=(2\pi)^{-1/2}\exp(-u^2/2)du$ and we have used 
$t=\sum\al (Q^{0})^{-1}\ab=1/(1+(n-1)q^p)\to 1/(1-q^p)$.
The condition that $\lambda_3$ is positive is given explicitly by
\be\label{RSAT}
 \dfrac{(1-q^p)^2}{2p(p-1)q^{p-2}(\epsilon+j_0m^p)^2} > 
 \int Du  \cosh^{-4}(\sqrt{\hat q}u + \hat m).
\ee
This is the stability condition for the replica symmetric solution
in the microcanonical ensemble.

Then, we compare the result with the canonical one.
The inverse temperature is defined as $\beta=ds/d\epsilon$ and 
is given in the RS case as $\beta=-2(\epsilon+j_0m^p)/(1-q^p)$. 
Inserting this relation into eq.~(\ref{RSAT}), 
we see that, as far as the replica symmetric solution is concerned, 
the microcanonical and canonical AT lines coincide formally.

\subsection{First-Step Replica Symmetry Breaking}

With only replica symmetry, 
the spin-glass phase does not show up for $p>2$, 
and we need to consider replica symmetry breaking. 
The 1RSB ansatz is characterized by $m\al=m$ and 
$q\ab=q_0+(q_1-q_0)\eta\ab(x)-q_1\delta\ab$, 
where $\eta\ab(x)$ is unity around the diagonal in blocks of size $x$ 
and zero else. 
In the absence of the external field, 
solution of the 1RSB saddle-point equations yields 
that the stable solution is always characterized by $q_0=0$ \cite{zbhn}. 
Subsequently, we will set $q_0=0$ and relabel $q_1=q$.

Proceeding along the same line as the RS case, 
we obtain the 1RSB AT condition as
\be\label{1rsb_at}
 \dfrac{(1-(1-x)q^p)^2}{2p(p-1)q^{p-2}(\epsilon+j_0m^p)^2} > 
 \dfrac{\int Du  [\cosh(\sqrt{\hat q}u+\hat m)]^{x-4}}
 {\int Du  [\cosh(\sqrt{\hat q}u+\hat m)]^{x}}.
\ee
Using the 1RSB energy-temperature relation 
$\beta=-2(\epsilon+j_0m^p)/(1-(1-x)q^p)$, 
we see that eq.~\eqref{1rsb_at} is equivalent to 
the canonical AT condition \cite{gardner}.

\section{Nishimori Line}\label{NL}

In this section, we show 
the ensemble equivalence of the NL.
In the original derivation of the NL in the microcanonical 
ensemble \cite{nml2}, the random average is denoted as 
\be
 s = C\prod_{\langle i_1\cdots i_p\rangle}\int 
 dJ_{\langle i_1\cdots i_p\rangle} P_0(J_{i_1\cdots i_p})
 \delta\left(N_{\rm B}j_0-\sum_{\langle i_1\cdots i_p\rangle}J_{i_1\cdots i_p}\right)
 \ln \Tr \delta (E-H),
 \label{j0nishimori}
\ee
 where $P_0(J)$ is given by eq.~(\ref{prob}) with $j_0=0$ and 
 $C$ is an irrelevant constant.
 This is different from the standard measure (\ref{prob}).
 The crucial difference is that the gauge invariance can be utilized 
 for eq.~(\ref{j0nishimori}) and not for eq.~(\ref{prob}).
 If we use eq.~(\ref{j0nishimori}), the microcanonical version of 
 the gauge invariant condition is given by $\epsilon=-j_0$.
 This corresponds to the condition for the canonical case 
 $\beta=2j_0$ \cite{nml1}, which implies that 
 the difference between their distributions is irrelevant. 
 We show explicitly this argument.

 If we use eq.~(\ref{j0nishimori}) for the distribution, 
 we obtain the same form as eq.~(\ref{S}) with the replacement
\be
 Q_{\alpha\beta} \to Q_{\alpha\beta}-m_{\alpha}^pm_{\beta}^p.
\ee
 However, this change does not affect the final form of the entropy.
 If we impose the 1RSB ansatz, we obtain the replacement 
\be
 \sum_{\alpha\beta}
 \epsilon_{\alpha}(Q)^{-1}_{\alpha\beta}\epsilon_{\alpha}
 &=& \frac{n}{1+(x-1)q^p}(\epsilon+j_0m^p)^2 \no\\
 &\to& \frac{n}{1+(x-1)q^p-nm^{2p}}(\epsilon+j_0m^p)^2.
\ee
 This difference disappears when we consider the $n\to0$ limit.
 Therefore, we conclude that 
 the difference between eqs.~(\ref{prob}) and (\ref{j0nishimori})
 is irrelevant and 
 we can utilize the gauge invariance 
 on the NL $\epsilon=-j_0$ in our model.

\section{Phase Diagram}\label{PD}

\begin{figure}[tbh]
\begin{center}
\subfigure[]{\includegraphics[width=0.45\textwidth]{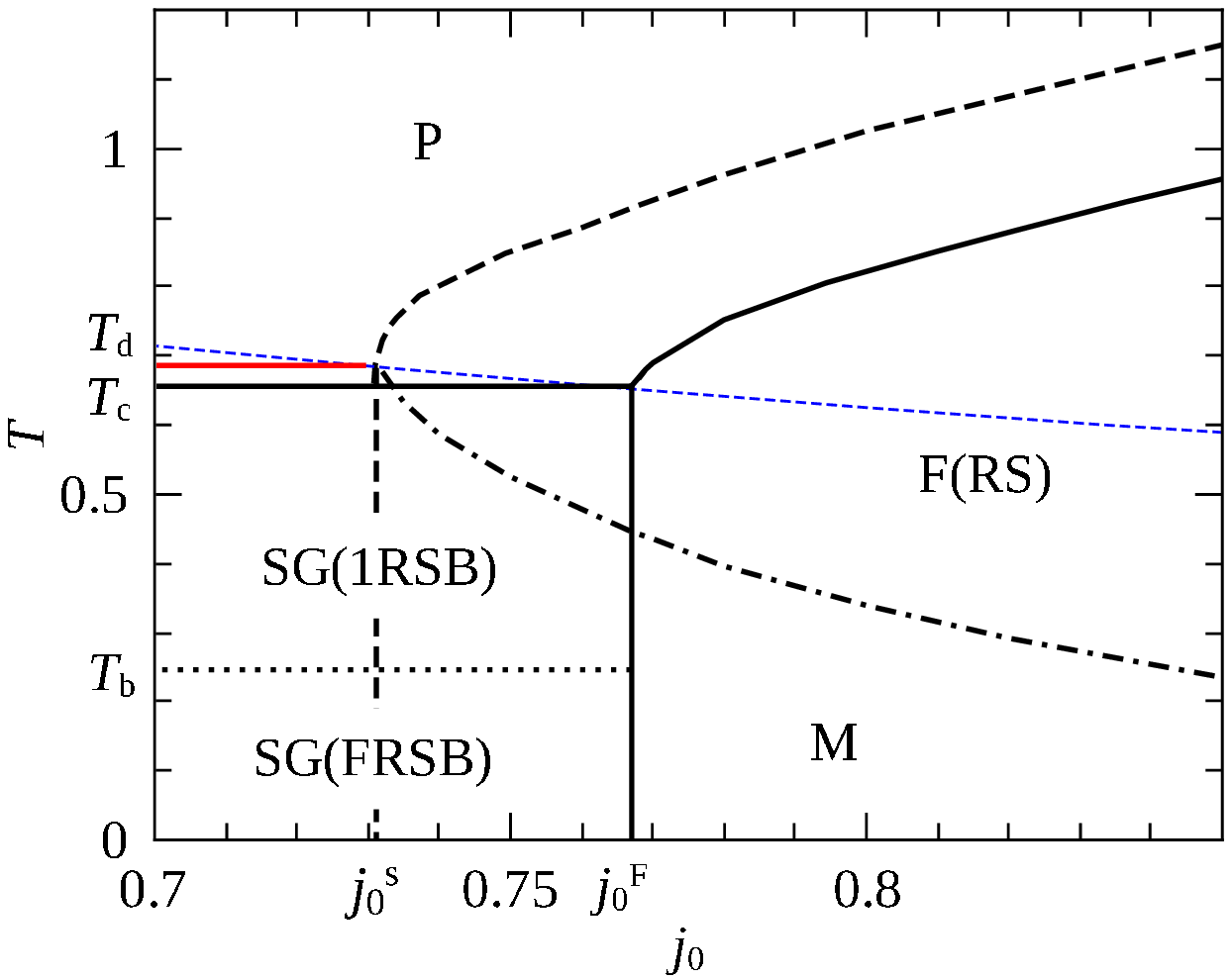}}
\subfigure[]{\includegraphics[width=0.45\textwidth]{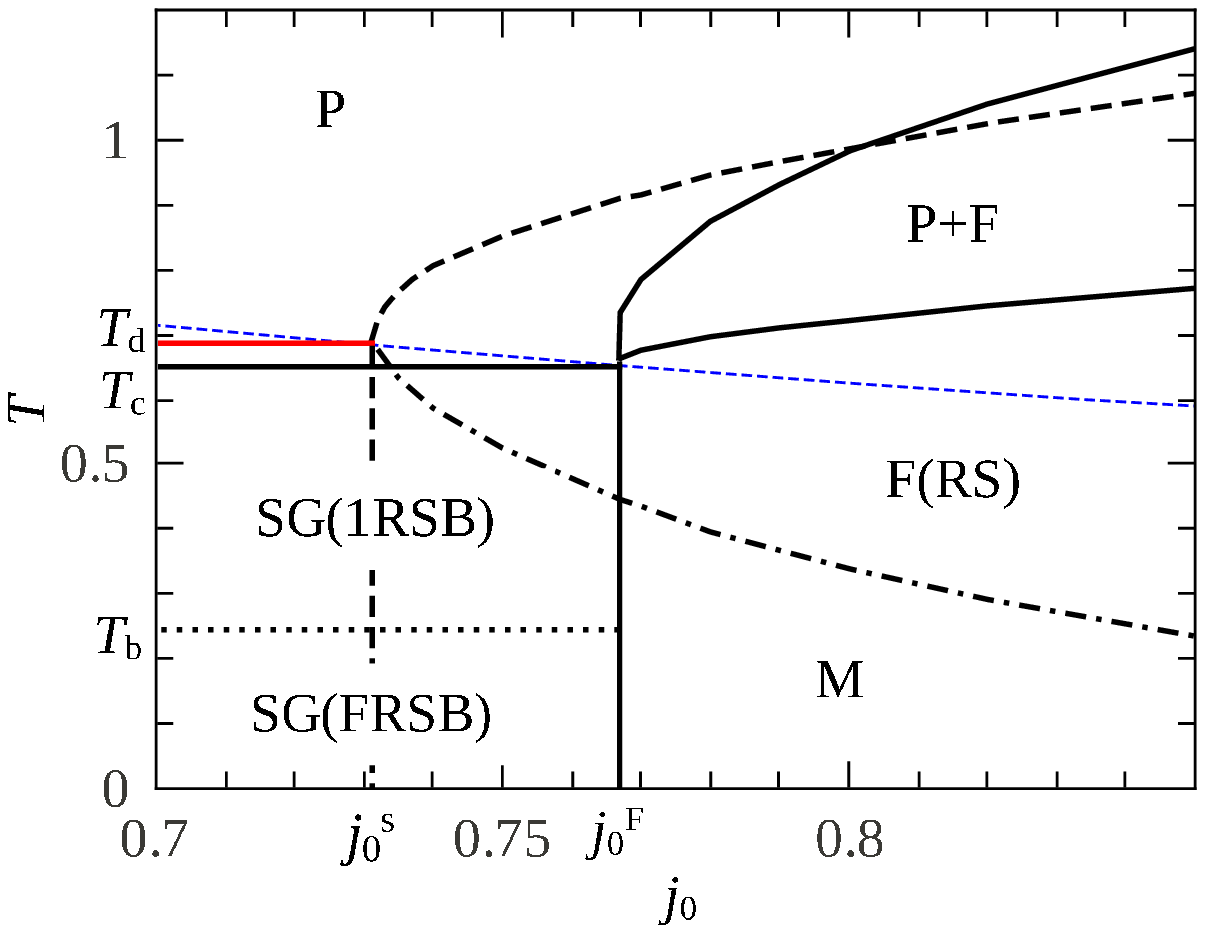}}
\end{center}
\caption{(Color online) Phase diagram of the many-body Ising spin glass with $p$=3 
in the a) canonical and b) microcanonical ensemble. 
The thermodynamic phase boundaries are drawn in solid black lines, 
while the dynamical transition is drawn in solid red. 
The limit of the metastability of the ferromagnetic phase (spinodal line) 
is drawn in black dashed, while the AT line, 
below which the replica-symmetric solution is unstable, 
is drawn black dash-dotted. 
The blue dashed line is the NL. 
The limit of the stability of the 1RSB solution of 
the spin-glass phase is drawn black dotted.}
\label{p3pd}
\end{figure}
\begin{figure}[hb]
\begin{center}
\includegraphics[width=0.45\textwidth]{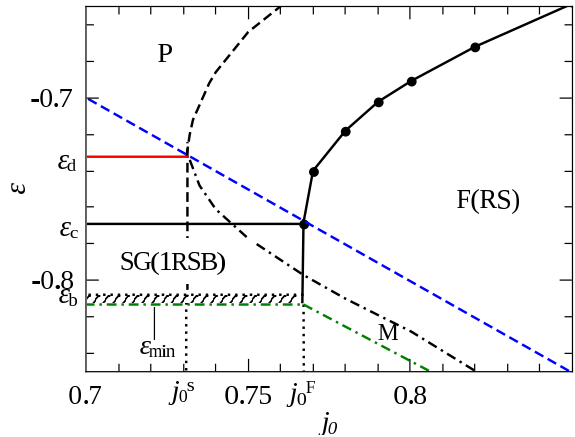} 
\end{center}
\caption{(Color online) Microcanonical phase diagram in the $(j_0,\epsilon)$-plane. 
The boundary between paramagnetic and ferromagnetic phases is drawn 
in black with circles. 
The spinodal line is drawn black dashed, 
while the AT line is shown black dash-dotted. 
The solid red line marks the dynamical transition. 
The full-RSB spin-glass phase exists only in the very narrow shaded region. 
The green dash-dotted line marks the minimal attainable energy. 
The blue dashed line is the NL.}
\label{p3ej}
\end{figure}

Having obtained the microcanonical AT line in the previous section, 
we are finally able to draw the complete microcanonical phase diagram. 
In fig.~\ref{p3pd} we compare the a) canonical and b) microcanonical 
phase diagrams. 
The canonical case was first obtained in \cite{nmwong} and 
the microcanonical phase diagram, except the AT line, 
was drawn in \cite{zbhn}. 
We re-draw here both for the sake of completeness. 
For $j_0<j_0^{\rm F}(\approx0.767)$ there is, in both ensembles, 
a horizontal, second-order phase boundary between a paramagnetic (P) 
and a 1RSB spin-glass (SG) phase at $T_c\approx 0.654$. 
The stability-boundary is ensemble-equivalent, i.e. the SG phase becomes 
unstable at $x\approx 0.335$, $T_{\rm b}\approx 0.245$ (black dotted), 
in the microcanonical as well as canonical ensembles. 
However, before the equilibrium P-SG transition takes place, 
there is a dynamical transition at $T=T_{\rm d}(\approx 0.686)$ (red), 
where the free energy develops an exponential number of minimal and 
the ergodicity breaks \cite{krzd}. 
The dynamical transition occurs at the same point in both ensembles. 
For $j_0>j_0^{\rm F}$, there exists a replica-symmetric 
ferromagnetic (F) phase, which is separated from the P phase by 
a first-order transition. 
In the canonical ensemble this is a simple line (black), 
while in the microcanonical ensemble there is a region of phase coexistence, P+F 
in between \cite{zbhn}. 
The ferromagnetic meta-stable states extend until $j_0^{\rm s}\approx 0.731$, 
and the spinodal lines are shown black dashed. 
In both ensembles the ferromagnetic, RS solutions become unstable 
below the AT line, shown as black dash-dotted. 
Below this line there is a mixed (M) phase, 
where there is ferromagnetic order as well as RSB. 
The AT line is the same in both ensembles. 
The NL, with $T=1/2j_0$, is shown blue dashed.

In fig.~\ref{p3ej} we show the microcanonical phase diagram 
in the ($j_0,\epsilon$)-plane. 
The F and P phases are separated, for $j_0>j_0^{\rm F}$, by a single line, 
shown black with circles. 
The replica symmetric F phase becomes unstable below the AT line, 
shown in black dash-dotted. 
The dynamical transition is at $\epsilon_{\rm d}\approx -0.732$ 
drawn in red. 
The condition for the NL reads $\epsilon=-j_0$ and is shown in blue dashed. 
The SG phase is stable for $j_0<j_0^{\rm F}$ between 
$\epsilon_{\rm c}(\approx -0.769)$ and $\epsilon_{\rm b}(\approx -0.809)$. 
To estimate the value of the minimal attainable energy, 
$\epsilon_{\rm min}$, shown green dash-dotted, of the system 
we can take the energy value where the 1RSB solution freezes. 
The energy where the entropy of the full RSB solution becomes zero at $T=0$, 
lies at a higher energy, due to a peculiarity of the replica trick. 
Namely, the requirement that entropy be minimal with respect to 
the spin-glass order parameter and the RSB-parameter. 
A suggestive reason for this occurrence is that the term in 
the entropy as given in eq.~(\ref{S})
\be
 \lim_{n\to 0}\frac{1}{n}\salb q\ab\hat q\ab 
 \underset{\rm 1RSB}{\longrightarrow}
 \lim_{n\to 0}\frac{1}{n}\left\{n^2q_0\hat q_0 
 +\frac{n}{x}x^2(q_1\hat q_1-q_0\hat q_0)-nq_1\hat q_1\right\}
\ee
changes sign at $n=1$ since $q_0<q_1$.

\begin{figure}[th]
\begin{center}
\includegraphics[width=0.45\textwidth]{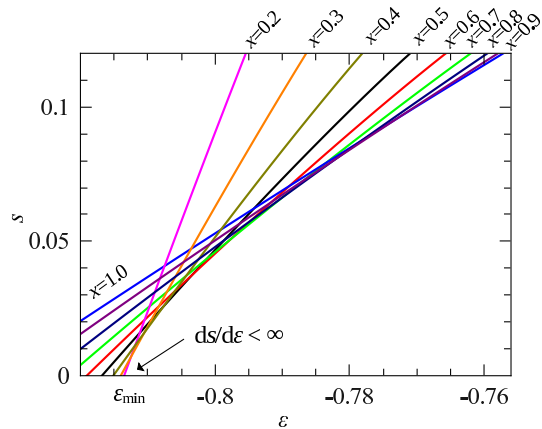} 
\end{center}
\caption{(Color online) Microcanonical 1RSB entropy at $j_0=0.5$ for various values of $x$. 
At the point where the entropy is zero, indicated by an arrow, 
its derivative is less than infinity. 
This results in a temperature larger than zero at this point.}
\label{sx}
\end{figure}

In fig.~\ref{sx} we show the entropy as a function of the energy 
for fixed values of $x$ at $j_0=0.5$. 
The equilibrium value of $s$ is determined 
by taking the lowest available entropy value. 
Comparing the curves at constant $x$, we see that 
the energy where the entropy is zero increases when $x$ decreases. 
For 1RSB, however, this procedure ends at $x\approx 0.2$. 
Lower values of $x$ do minimize the entropy but lead to negative values. 
We see furthermore that the energy where the entropy and 
temperature are both zero simultaneously when considering full RSB 
must lie at higher energy than $\epsilon_{\rm min}$. 
A similar statement holds for $j_0>j_0^{\rm F}$, for the ferromagnetic phase: 
The minimal attainable energy of the full RSB
solution is expected to lie above $\epsilon_{\rm min}$ which is the minimal energy of the 1RSB solution.

Between $\epsilon_{\rm min}$ and the AT line there exists again, for $j_0>j_0^{\rm F}$, 
a phase M which shows both, ferromagnetic order and RSB.
On the other hand, 
the maximal attainable energy of the system is at $\epsilon=0$.

\section{Conclusion}\label{CC}
In conclusion, we have shown that for Ising spin glasses 
with many-body interactions the AT condition yields the same curve 
in the canonical and microcanonical ensembles. 
Our significant result show that there is no ensemble inequivalence 
on the AT line. 
Since the ensembles are equivalent on NL as well and the AT line 
lies strictly at lower temperatures than the NL, 
we can surmise that there is no ensemble inequivalence below the AT line. 
This hypothesis is supported by another recent result \cite{bt} 
for spin glasses with integer spins, 
where the spin-glass phase transition can be ensemble inequivalent. 
There the AT line, however, terminates well before 
there is ensemble inequivalence.

\section*{Acknowledgements}
We thank Y. Matsuda and H. Nishimori for pointing out 
various initial mistakes and other valuable comments.



\begin{thebibliography}{13}
\bibitem{EA}  S. F. Edwards and P. W. Anderson: J. Phys. F \textbf{5} (1975) 965.
\bibitem{AT} J. R. L. de Almeida and D. J. Thouless: J. Phys. A \textbf{11} (1978) 983.
\bibitem{SK} D. Sherrington and S. Kirkpatrick: Phys. Rev. Lett. \textbf{35} (1975) 1792.
\bibitem{gardner} E. Gardner: Nucl. Phys. B \textbf{257} (1985) 747.
\bibitem{zbhn} Z. Bertalan and H. Nishimori: Phil. Mag. \textbf{92} (2012) 2.
\bibitem{campa} A. Campa, T. Dauxois and S. Ruffo: Phys. Rep. \textbf{480} (2009) 57.
\bibitem{nml1} H. Nishimori: Prog. Theor. Phys. \textbf{66} (1981) 1169.
\bibitem{nml2} H. Nishimori: J. Phys. Soc. Jpn. \textbf{80} (2011) 023002.
\bibitem{nishimoribook} H. Nishimori: \textit{Statistical Physics of Spin Glasses and Information Processing: An Introduction} (Oxford University Press, Oxford, 2001).
\bibitem{nmwong} H. Nishimori and K. Y. M. Wong: Phys. Rev. E \textbf{60} (1999) 132.
\bibitem{krzd} F. Krzakala and L. Zdeborova: J. Chem. Phys.  \textbf{134} (2011) 034512; 
 F. Krzakala and L. Zdeborova : J. Chem. Phys.  \textbf{134} (2011) 034513.
\bibitem{bt} Z. Bertalan and K. Takahashi: J. Stat. Mech. (2011) P11022.

\end{thebibliography}
\end{document}